\documentclass[prd,amsmath,amssymb,superscriptaddress,nofootinbib]{revtex4}
\usepackage{mathrsfs}
\usepackage{graphicx,subfigure}
\usepackage{hyperref}
\usepackage{psfrag}
\usepackage{color}

\newcommand{\scrim}{{$\mathscr{I}^-$}}
\newcommand{\hp}{{$\mathscr{H}^+$}}
\newcommand{\hm}{{$\mathscr{H}^-$}}

\begin{document}

\title{On the impossibility of superluminal travel: the warp drive lesson}

\date{\today}

\author{Carlos Barcel\'o}
\email{carlos@iaa.es}
\affiliation{Instituto de Astrof\'{\i}sica de Andaluc\'{\i}a, CSIC, Camino Bajo de Hu\'etor 50, 18008 Granada, Spain}
\author{Stefano Finazzi}
\email{finazzi@sissa.it}
\author{Stefano Liberati}
\email{liberati@sissa.it}
\affiliation{SISSA, via Beirut 2-4, Trieste 34151, Italy;\\  INFN sezione di Trieste}

\begin{abstract}

The question of whether it is possible or not to surpass the speed of light is already centennial. The special theory of relativity took the existence of a speed limit as a principle, the light postulate, which has proven to be enormously predictive. Here we discuss some of its twists and turns when general relativity and quantum mechanics come into play. In particular, we discuss one of the most interesting proposals for faster than light travel: warp drives. Even if one succeeded in creating such spacetime structures, it would be still necessary to check whether they would survive to the switching on of quantum matter effects. Here, we show that the quantum  back-reaction to warp-drive geometries, created out of an initially flat spacetime, inevitably lead to their destabilization whenever superluminal speeds are attained. We close this investigation speculating the possible significance of this further success of the speed of light postulate.

\end{abstract}


\maketitle

\section{A question for a Century}\label{sec:intro}

{\em Why is it not possible to travel faster than light?} Probably this is the most-frequently-asked question to scientists around the globe during the last century. Science does not have yet a compelling answer to this question and, logically possible but improbable in practice, it might even be that there is none. The speed of light as a maximum speed for propagation of any signal, the light postulate, was introduced by Einstein as a hypothesis or principle in his famous paper of 1905. Einstein himself acknowledged that his theory of Special Relativity (SR) was a ``principle theory" to be validated empirically and not a ``constructive theory" trying to explain the facts from elementary foundations~\cite{brown-timpson,RBrown:2005p490}. 

After the proposal of the relativity principle together with the light postulate, a large part of the developments in physics during the last Century came to life from the desire of making all theories compatible with these principles. From a predictive point of view, these principles have been a tremendously successful source of inspiration in physics and up to now, there is not a single observation contradicting the light postulate.

On another front, the exploration of the universe has enlarged further and further its size to inconceivable proportions. Given the current way in which we humans understand this exploration, that is, remaining {  on} the Earth while sending round-trip expeditions outside, it is almost unavoidable not to feel from time to time that the speed of light barrier restrain our probing capacities to unbearable limits. That is one of the reasons why, from time to time, scientists like to revise the relativistic scientific building to look for fissures. In most of the cases the consistency fissures will ask for healing conditions which, if empirically correct, will add more medals to the impressive trophy shelf of the relativity principle and the light postulate. But one cannot discard that through some of these fissures one might glimpse an extended theory allowing for superluminal travel. In any case, pushing physics to its limits has always been a source of advancement and in this essay we will give a recount of one particular battle on the fields of the light postulate we have participate in.

The general theory of relativity (GR) was born from the desire of constructing a gravitational theory consistent with the relativity principle and the light postulate. At a first glance general relativity does incorporate the light postulate, but in a subtle and somewhat restricted manner: No signal can travel faster than the speed of light as defined locally with respect to space and time, or in other words, the spacetime geometry is everywhere Lorentzian. General relativity tell us that gravity is encoded in terms of Lorentzian geometry. However, although this assertion has an enormous significance, it does not say anything about our real chances of sending an expedition to our neighboring star, Alpha Centauri, and receiving it back in less that 8.6 years.

\section{A geometry for superluminal travel: The warp drive}\label{sec:suplumtravel}

\begin{figure}
\includegraphics[width=.7\textwidth]{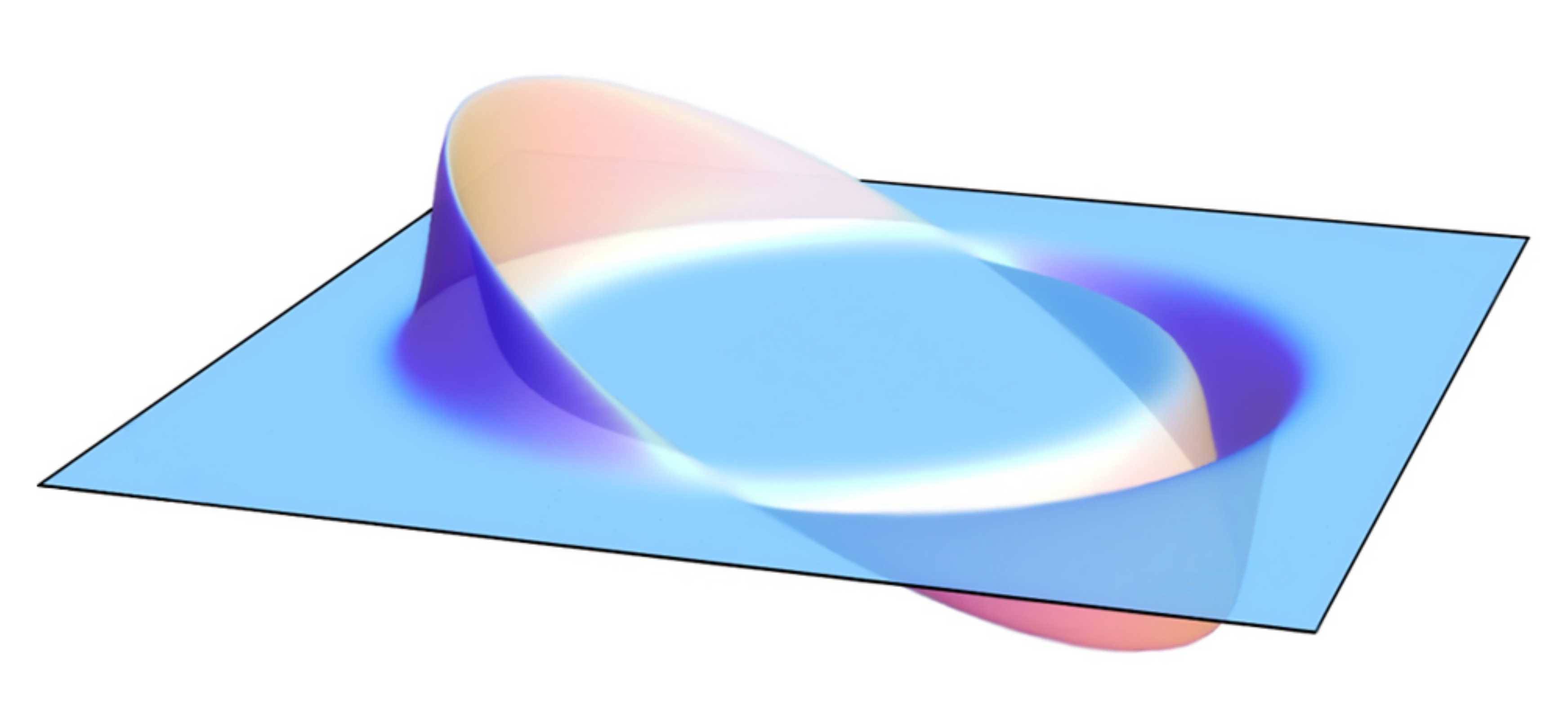}
 \caption{Spacetime structure of a warp-drive bubble. (Source: internet)}
 \label{fig:wd-stru}
\end{figure}

Nothing can travel faster than light with respect to space, but what about space itself? The kinematics of GR sets no restriction on the expanding or contracting capacities of spacetime itself. By manipulating the light-cone structure of Minkowski spacetime one can construct geometries allowing for superluminal travel. A prime example of that is the warp-drive geometry introduced by Miguel Alcubierre in 1994~\cite{alcubierre}. This geometry represents a bubble containing an almost flat region, moving at arbitrary speed within an asymptotically flat spacetime. Mathematically its metric can be written as
\begin{equation}\label{eq:3Dalcubierre}
ds^2=-c^2 dt^2+\left[dx-v(r)dt\right]^2+dy^2+dz^2~,
\end{equation}
where $r\equiv \sqrt{[x-x_c(t)]^2+y^2+z^2}$ is the distance from the center of the bubble, $\{x_c(t),0,0\}$, which is moving in the $x$ direction with arbitrary speed $v_c=dx_c/dt$. Here $v(r)=v_c f(r)$ {  and} $f$ is a suitable smooth function satisfying $f(0)=1$ and $f(r) \to 0$ for $r\to \infty$.  In Fig.~\ref{fig:wd-stru}, the curvature of the warp-drive geometry is plotted: To make the warp-drive travel at the speed $v_c(t)$, the spacetime has to contract in front of the warp-drive bubble and expand behind. It is easy to see that the worldline $\{x_c(t),0,0\}$ is a geodesic for the above metric. Roughly speaking, if one places a spaceship at $\{x_c(t),0,0\}$, it is not subject to any acceleration, while moving faster than light with respect to someone living outside of the bubble. 

Looking at the previous geometry it would seem that general relativity easily allows superluminal travel; but this is not quite true. General relativity is not only Lorentzian geometry, in addition one has to carefully specify the right hand side of the Einstein equations, that is, the stress-energy tensor of the matter content. When the warp-drive geometry is interpreted as a solution of the Einstein equations one realizes that the matter content supporting it has to be ``exotic", {\em i.e.} it has to violate the so called energy conditions (EC) of GR~\cite{hawking-ellis,visser-book}.

%

\section{The attractive character of gravity}\label{sec:gravity}

Our daily experience tell us that gravity is attractive  or, what is equivalent, that the mass (energy) of a body is always positive. The energy conditions of GR mathematically encode this observation. They take the form of inequalities involving the full stress energy tensor of matter (both {  energy density and pressure} gravitate in GR) which determines the focussing/defocussing properties of the gravitational field via the Einstein equations. Hence, the EC impose restrictions on the allowed manipulations of {  light cones} (i.e. on the local causal structure of spacetime). 

Indeed, it seems that any attempt to produce superluminal travel would need some matter with gravitationally repulsive properties~\cite{olum,visser-censorship}.  In particular, as we anticipated above, the light cone structure of the Alcubierre warp drive requires violations of the weak and dominant energy conditions~\cite{pfenningford} (remarkably this is true even if the warp drive is not traveling at superluminal speeds). 

One could say that any strong version of the light postulate, forbidding superluminal travel of any sort and not just of the local type, will be linked to the gravitationally attractive properties of matter. In fact, not only the Alcubierre warp drive but also alternative ``spacetime shortcuts", such as the Krasnikov tube~\cite{1998PhRvD..57.4760K, everett-roman} or traversable wormholes~\cite{visser-book}, seem to require the same kind of exotic matter~\cite{lobo2007}. Of course, empirical observations in the future will decide whether a strong light postulate is at work or not. At present{,} theoretical investigations and empirical evidence are still not completely in favor of its existence. 

On the one hand, energy conditions can be violated by several physical systems, even classical ones~\cite{Barcelo:2002bv}. 
On the other hand, violations of the weak and dominant energy conditions are particularly difficult to get as they imply negative energy densities. Quantum phenomena, such as the Casimir effect, are known to entail such violations~\cite{visser-book} (and indeed it is a subject of debate their relevance for possible faster than light propagation~\cite{Scharnhorst:1990sr, Barton:1992pq, Liberati:2001sd, Liberati:2000mp}). However, it has been convincingly argued that these quantum mechanical violations of the energy conditions would have to satisfy (by the very same tents of Quantum Mechanics) strict bounds on their extension in time and space. These bounds are the so called \emph{quantum inequalities} (QI)~\cite{roman}. 

Hence, it is not so surprising that{,} immediately after Alcubierre's proposal of the warp drive, the most investigated aspect of its solution has been the amount of exotic matter required to support such a spacetime~\cite{pfenningford,broeck,lobovisser2004a,lobovisser2004b}. Applying QI  to the warp drive it has been found that such exotic matter must be confined in Planck-size regions at the edges of the bubble~\cite{pfenningford}, thus making the bubble-wall thickness to be of the order of the Planck length, $L_{\rm P}\approx  10^{-35}$m (see Fig.~\ref{fig:wd-en}). This bound on the wall thickness turns into lower limits on the amount of exotic matter required to support the bubble (at least of the order of 1 solar mass for a macroscopic bubble traveling at the speed of light). 

The requirement of exotic matter in order to support the warp drive can be seen as an engineering problem. Let us assume that some advance civilization would be finally able to solve it. Even in this case there would be another important issue regarding the feasibility of the warp drive: its semiclassical stability. This will be the subject of our investigation.

\section{On curvature and vacuum fluctuations}\label{sec:vacuumfluctuation}

In quantum field theory (QFT) the vacuum state possesses, at least formally, an infinite amount of energy (it can be understood as an infinite collection of harmonic oscillators, each contributing with energy $\hbar \omega/2$). However, to date, non-gravitational particle-physics phenomena seem to depend only on energy differences between states, so the value of the quantum vacuum energy does not play any role: The vacuum contribution to the total {stress-energy} tensor (SET) of any field in flat spacetime is renormalized to zero using a subtraction scheme. In a curved spacetime the divergent part of the SET can still be canceled, by using the same subtraction scheme that works in flat spacetime. However, the subtraction is now no longer exact, leaving a finite residual value for the renormalized SET (RSET) --- this effect is called quantum vacuum polarization. 

Therefore, one ends up with the following iterative process: Classical matter curves spacetime via Einstein equations, by an amount determined by its classical SET; this curvature makes the quantum vacuum acquire a finite non-vanishing RSET; the latter becomes an additional source of gravity, modifying the initial curvature; the new curvature induces in turn a different RSET, and so on. In this way one can incorporates quantum corrections into General Relativity in a ``minimal'' way, taking into account the quantum behavior of matter but still treating gravity (that is, spacetime) classically. Hence, the name ``semiclassical approach".

\begin{figure}
\includegraphics[width=.5\textwidth]{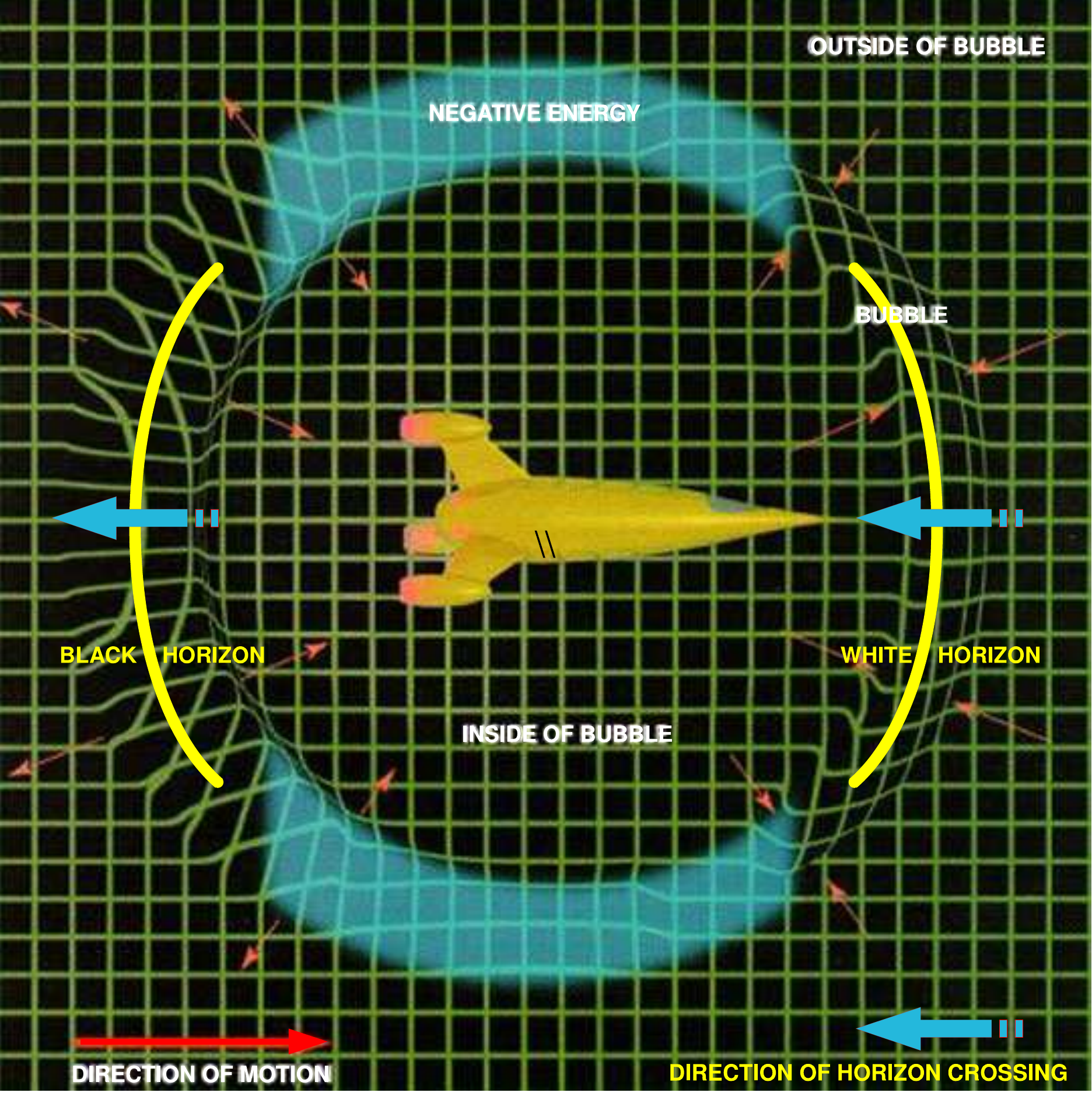}
 \caption{Artistic representation of a Warp Drive.  From the point of view of an observer inside the bubble, the front (back) wall looks like the horizon of a white (black) hole (\emph{yellow solid lines}).  Large amounts of exotic matter are concentrated in the walls on a plane orthogonal to the direction of motion. (Source: {\it Scientific 
American})}
 \label{fig:wd-en}
\end{figure}

The stability of the stationary (eternal) superluminal warp-drive geometry against the quantum-vacuum effects was studied in~\cite{hiscock}. There it was noticed that, to an observer within the warp-drive bubble, the backward and forward walls (along the direction of motion) look respectively as the future (black) and past (white) event horizon of a black hole (see Fig.~\ref{fig:wd-en}). We name them respectively the black and white horizon of the bubble. Indeed, while the warp-drive bubble travels at superluminal speeds, nothing can escape from inside the bubble to the external world passing through the front wall, neither anything can enter the bubble from the back, as this would require that signals travel locally faster than light (remember that in GR special relativity still rules locally).

In this essay we consider the realistic case of a warp drive created with zero velocity at early times and then accelerated up to some superluminal speed in a finite time (a more detailed treatment can be found in \cite{wd}). Spacetime in the past is flat, therefore the physical vacuum state has to match the Minkowski vacuum at early times (we work in the Heisenberg representation). At late times we find, as expected, that the center of the bubble is filled with a thermal flux of radiation at the Hawking temperature corresponding to the surface gravity of the black horizon. The latter is inversely proportional to the wall thickness. Hence, if the QI hold, then Planck-size walls would lead to an excruciating temperature of the order of the Planck temperature $T_{\rm P}$ ($10^{32}$ in the Kelvin/absolute scale or in whatever temperature scale one adopts!). Even worse, we do show that the RSET does increase exponentially with time on the white horizon (while it is regular on the black one). This clearly implies that a warp drive becomes rapidly unstable once a superluminal speed is reached. You may be able to build a warp drive but you still will have to respect the speed of light limit.

\section{Scheme of the calculation}\label{sec:calculation}

We are including now a technical section with the scheme of our calculation. The reader not interested in mathematical details can safely ``jump'' directly to Sect.~\ref{sec:results}, where the results are briefly summarized.

\subsection{Light-ray propagation}

In the actual computation we shall restrict our attention to the $1+1$ dimensions case (since this is the only one for which one can carry out a complete analytic treatment as explained below).\footnote{Indeed, we do expect that the salient features of our results would be maintained in a full 3+1 calculation, given that they will still be valid in a suitable open set of the horizons centered around the axis aligned with the direction of motion.}
Changing coordinates to those associated with an observer at the center of the bubble, the warp-drive metric~\eqref{eq:3Dalcubierre} becomes
\begin{equation}\label{eq:fluidmetric}
 ds^2=-c^2 dt^2+\left[dr-\bar{v}(r)dt\right]^2,\qquad \bar{v}=v-v_c~,
\end{equation}
where $r\equiv x-x_c(t)$ is now the signed distance from the center of the bubble. In our dynamical situation the warp-drive geometry interpolates between an initial Minkowski spacetime [$\hat{v}(t,r)\to0$, for $t \to -\infty$] and a final stationary superluminal ($v_c>c$) bubble [$\hat{v}(t,r)\to\bar{v}(r)$, for $t \to +\infty$]. To an observer living inside the bubble this geometry has two horizons, a \emph{black horizon} \hp~located at $r_1$ and a \emph{white horizon} \hm~located at $r_2$. For those interested, in~\cite{wd} you can find the Penrose diagram of these spacetimes.
Here let us just mention that from the point of view of the Cauchy development of \scrim~these spacetimes posses Cauchy horizons.

\begin{figure}[tb]
\includegraphics[width=0.4\textwidth]{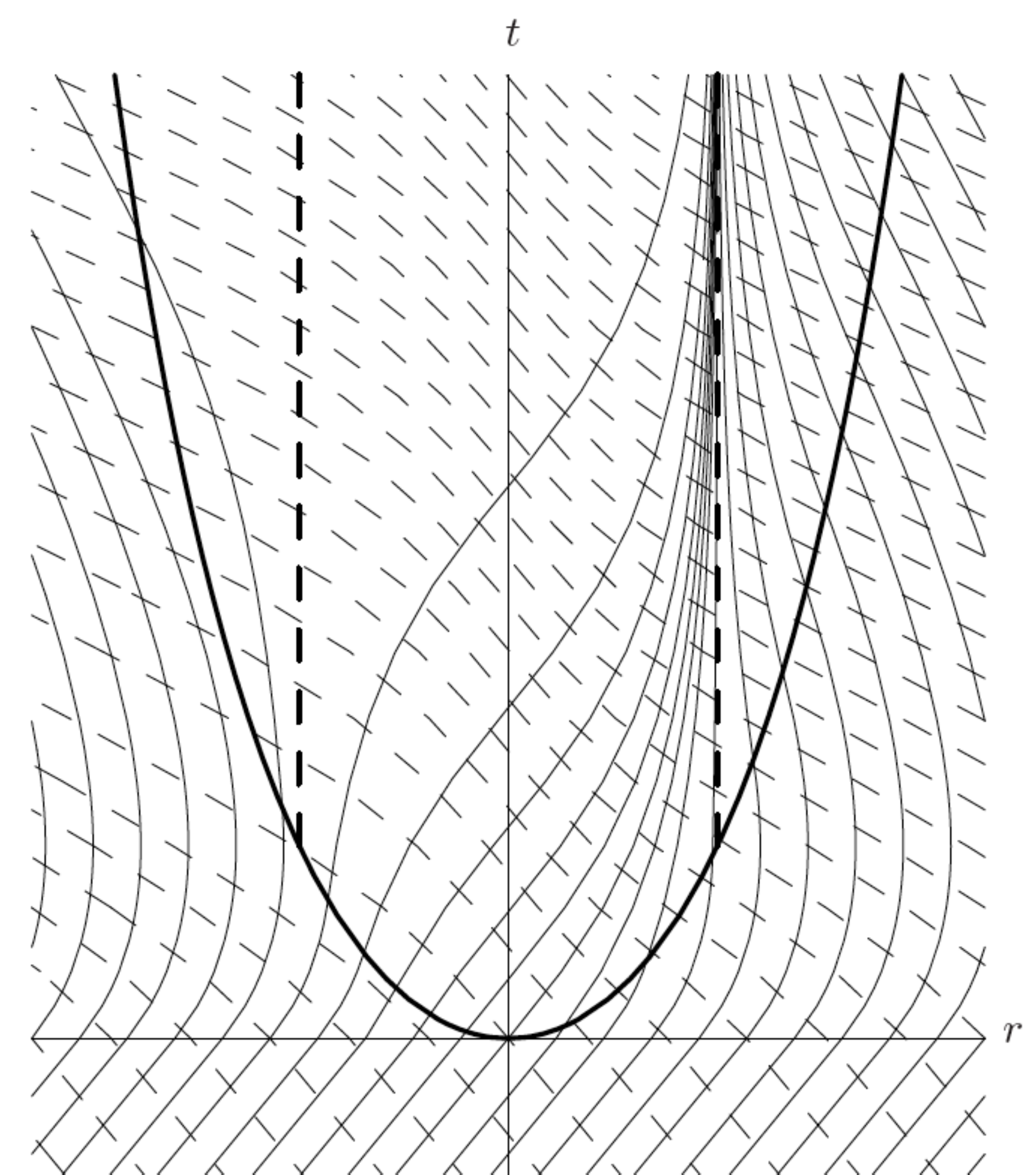}
\caption{Light rays propagating rightward (\emph{solid lines}) and leftward (\emph{dashed lines}) in the plane $(t,r)$ in a warp-drive spacetime. At $t<0$ the metric is Minkowskian. The horizons at $r_1$ and $r_2$ (\emph{heavy dashed lines}) are formed at $t=T_H=1$.}
\label{fig:lightrayssimp}
\end{figure}

Let us now consider light-ray propagation in the above described geometry. Only the behavior of right-going rays determines the universal features of the RSET, just like outgoing modes do in the case of a black hole collapse \cite{wd,particlecreation,stresstensor}. Therefore, we need essentially the relation between the past and future null coordinates $U$ and $u$, labelling right-going light rays (see {  Fig.}~\ref{fig:lightrayssimp}). Following \cite{particlecreation}, {  this relation can be found by integrating the {  right-going-ray} equation} 
\begin{equation}\label{eq:rays}
 \frac{dr}{dt}=c+\hat{v}(r,t)~.
\end{equation}

{  There are two special right-going rays defining, respectively, the asymptotic location of the black and white horizons. In terms of the right-going past null coordinate $U$ let us denote these two rays by $U_{\rm BH}$ and $U_{\rm WH}$, respectively. The finite interval $U \in (U_{\rm WH}, U_{\rm BH})$ is mapped to the infinite interval $u \in (-\infty, +\infty)$ covering all the rays traveling inside the bubble. For rays which are close to the black horizon, in~\cite{wd} the present authors proved that the relation between $U$ and $u$ can be approximated as a series of the form}
\begin{eqnarray}\label{eq:Uu}
U(u\to +\infty) \simeq U_{\rm BH} + A_1 e^{-\kappa_1 u} + \frac{A_2}{2}e^{-2\kappa_1 u} + \dots ~.
\end{eqnarray}
Here $A_n$ are constants (with $A_1<0$) and $\kappa_1>0$ represents the surface gravity of the black horizon. This relation is the standard result for the formation of a black hole through gravitational collapse. As a consequence, the quantum state which is vacuum on \scrim~will show, for an observer inside the warp-drive bubble, Hawking radiation with temperature $T_{H}=\kappa_1 /2\pi$.

{Equivalently, we find that the corresponding expansion in proximity of the white horizon is} 
\begin{eqnarray}
U(u\to -\infty) \simeq U_{\rm WH} + D_1 e^{\kappa_2 u} + \frac{D_2}{2}e^{2\kappa_2 u} + \dots ~,
\end{eqnarray}
where $D_2>0$ and $\kappa_2$ is the white hole surface gravity and is also defined to be positive ($\kappa_2$ could be different from $\kappa_1$ in general, although it is expected to be comparable with $\kappa_1$ in this specific case). The interpretation of this relation in terms of particle production is not as clear as in the black horizon case. For this reason, we shall consider now the RSET.

\subsection{Renormalized stress-energy tensor}

For the calculation of the RSET inside the warp-drive bubble we use the method proposed in~\cite{stresstensor}. In past null coordinates $U$ and $W$ the metric can be written as
\begin{equation}\label{eq:metricUW}
 ds^2=-C(U,W)dUdW~.
\end{equation}
In the {stationary region {at late times}, we use the previous future null coordinate $u$} and $\tilde w$, defined as 
\begin{eqnarray}\label{eq:wdef}
\tilde w(t,r)= t + \int_{0}^{r} \frac{dr}{c-\bar{v}(r)}~.
\end{eqnarray}
In these coordinates the metric is expressed as 
\begin{equation}\label{eq:metricuw}
 ds^2=-\bar{C}(u,\tilde w) du d\tilde w~, \quad C(U,W) = \frac{\bar{C}(u,\tilde w)}{\dot{p}(u)\dot{q}(\tilde w)}~,
\end{equation}
where 
$U=p(u)$ and $W=q(\tilde w)$. In this way, $\bar{C}$ depends only on $r$ through $u,\tilde w$.

For concreteness, we refer to the RSET associated with a quantum massless scalar field living on the 
spacetime. The RSET components have the following form \cite{birreldavies}:
\begin{align}
 T_{UU} &= -\frac{1}{12\pi}C^{1/2}\partial_U^2 C^{-1/2}~,\label{eq:TUU}\\
 T_{WW} &= -\frac{1}{12\pi}C^{1/2}\partial_W^2 C^{-1/2}~,\label{eq:TWW}\\
 T_{UW} &= T_{WU} =\frac{1}{96\pi}C~R~.\label{eq:TUWR}
\end{align}
If there were other fields present in the theory, the previous expressions would be multiplied by a specific numerical factor. Using the relationships $U=p(u)$, $W=q(\tilde w$) and the time-independence of $u$ and $\tilde w$, one can calculate~\cite{wd} the RSET components in the stationary region:
\begin{align}
 T_{UU} &= -\frac{1}{48\pi}\frac{1}{\dot{p}^2}\left[\bar{v}'\,^2+\left(1-\bar{v}^2\right)\bar{v}\bar{v}''-f(u)\right]~,\\
 T_{WW} &= -\frac{1}{48\pi}\frac{1}{\dot{q}^2}\left[\bar{v}'\,^2+\left(1-\bar{v}^2\right)\bar{v}\bar{v}''-g(\tilde w)\right]~,\\
 T_{UW} &= T_{WU}=-\frac{1}{48\pi}\frac{1}{\dot{p}\dot{q}}\left(1-\bar{v}^2\right)\left[\bar{v}'\,^2+\bar{v}\bar{v}''\right]~,
\end{align}
where we have put $c=1$ and we have defined
\begin{align}
 f(u)&\equiv\frac{3\ddot{p}^2(u)-2\dot{p}(u)\,\dddot{p}(u)}{\dot{p}^2(u)}~,\label{fu}\\
 g(\tilde w)&\equiv\frac{3\ddot{q}^2(\tilde w)-2\dot{q}(\tilde w)\,\dddot{q}(\tilde w)}{\dot{q}^2(\tilde w)}~.
\end{align}
One can show~\cite{wd} that $\dot{q}$ contains solely information associated with the dynamical details of the transition region. Moreover, for simple dynamical interpolations between Minkowski and the final warp drive, $\dot{q}(\tilde w )$ goes to a constant at late times, such that $g(\tilde w)\to 0$. From now on, we will neglect this term.

We want to look at the energy density inside the bubble, in particular at the energy $\rho$ as measured by a set of free-falling observers, whose four velocity is $u_{c}^\mu=(1,\bar{v})$ in $(t,r)$ components. For these observers we obtain
\begin{equation}
 \rho=T_{\mu\nu}u_{c}^\mu u_{c}^\nu=\rho_{\rm st}+\rho_{{\rm dyn}}~,\\
\end{equation}
where we define a static term $\rho_{\rm st}$, depending only on the $r$ coordinate through $\bar{v}(r)$,
\begin{equation}
 \rho_{\rm st}\equiv-\frac{1}{24\pi}\left[\frac{\left(\bar{v}^4-\bar{v}^2+2\right)}{\left(1-\bar{v}^2\right)^2}\bar{v}'\,^2+\frac{2\bar{v}}{1-\bar{v}^2}\bar{v}''\right]~,\label{eq:rhocs}
 \end{equation}
  and a dynamic term $\rho_{{\rm dyn}}$
  \begin{equation}
 \rho_{{\rm dyn}}\equiv\frac{1}{48\pi}\frac{f(u)}{\left(1+\bar{v}\right)^2}~.\label{eq:rhodu}
\end{equation}
These latter term, depending also on $u$, corresponds to energy travelling on right-going rays, eventually red/blue-shifted by a term depending on $r$.

\subsection{Hawking radiation inside the bubble}

We study now the behavior of the RSET in the center of the bubble at late times. Here $\rho_{\rm st}=0$, because $\bar{v}(r=0)=\bar{v}'(r=0)=0$. {  Integrating Eq.~\eqref{eq:rays},} one realizes that $u(t,r)$ is linear in $t$ so that, for fixed $r$, it acquires with time arbitrarily large positive values. One can evaluate $\rho_{\rm dyn}$ from Eq.~\eqref{eq:rhodu} by using a late-time expansion for $f(u)$, which (up to the first non-vanishing order in $e^{-\kappa_1 u}$) gives $f(u) \approx \kappa_1^2$, so that $\rho(r=0) \approx \kappa_1^2/(48\pi)=\pi T_H^2/12$, where $T_H \equiv \kappa_1/(2\pi)$ is the usual Hawking temperature. The above expression is the energy density of a scalar field in $1+1$ dimension at finite temperature $T_H$. This result confirms that an observer inside the bubble measures a thermal flux of radiation at temperature $T_H$.

\subsection{Problems with horizons}

Let us now study $\rho$ on the horizons \hp~and \hm. Here, both $\rho_{\rm st}$ and $\rho_{{\rm dyn}}$ are divergent because of the $(1+\bar{v})$ factors in the denominators. Using the late time expansion of $f(u)$ in the proximity of the black horizon~\cite{wd}
\begin{equation}
\lim_{r\to r_1 }f(u) = \kappa_1^2\left\{1+\left[3{\left(\frac{A_{2}}{A_{1}}\right)}^2-2\frac{A_{3}}{A_{1}}\right] e^{-2\kappa_1 t}\left(r-r_1\right)^2+{\cal O}\left(\left(r-r_1\right)^3\right) \right\}~,
\end{equation}
and expanding both the static and the dynamic {terms} up to order ${\cal O}(r-r_1)$, one obtains that the diverging terms ($\propto(r-r_1)^{-2}$ and $\propto(r-r_1)^{-1}$) in $\rho_{\rm st}$ and $\rho_{{\rm dyn}}$ exactly cancel each other~\cite{wd}. An analogous cancellation is found when studying the formation of a black hole through gravitational collapse~\cite{stresstensor}. It is now clear that the total $\rho$ is ${\cal O}(1)$ on the horizon and does not diverge at any finite time (as expected from Fulling-Sweeny-Wald theorem \cite{fsw}).
By looking at the subleading terms,
\begin{equation}
 \rho=\frac{e^{-2\kappa_1 t}}{48\pi}\left[3{\left(\frac{A_2}{A_1}\right)}^2-2\frac{A_3}{A_1}\right] + A
	+{\cal O}\left(r-r_1\right)~,
\end{equation}
where $A$ is a constant, we see that on the black horizon the contribution of the transient radiation (different from Hawking radiation) dies off {  exponentially with time, on} a time scale $\sim 1/\kappa_1$.\footnote{However, in analogy to the conclusions of~\cite{stresstensor}, a slow approach to the black-horizon formation might lead to large values of the RSET and hence to a large back-reaction.}

Close to the white horizon, the divergences in the static and dynamical contributions cancel each other, as in the black horizon case. However, something distinctive occurs with the subleading contributions. In fact, they now becomes
\begin{equation}
 \rho=\frac{e^{2\kappa_2 t}}{48\pi}\left[3{\left(\frac{D_2}{D_1}\right)}^2-2\frac{D_3}{D_1}\right] + D
	+{\cal O}\left(r-r_1\right)~.
\end{equation}
This expression shows an exponential increase of the energy density with time. This means that {  $\rho$ grows exponentially and eventually diverges along \hm}.

In a completely analogous way, one can study $\rho$ close to {  the Cauchy horizon}~\cite{wd}. Performing an expansion at late times ($t\to+\infty$) one finds that the RSET diverges also there, without any contradiction with the Fulling-Sweeny-Wald theorem~\cite{fsw}, because this is precisely a Cauchy horizon.

Note that the above mentioned divergences are very different in nature. The divergence at {late times on \hm}~stems from the untamed growth of the transient disturbances produced by the white horizon formation. The RSET divergence on {the Cauchy horizon} is due instead  to the well known infinite blue-shift suffered by light rays while approaching this kind of horizon. It is analogous to the often claimed instability of inner horizons in Kerr-Newman black holes~\cite{Simpson:1973ua,poissonisrael,markovicpoisson}.
Anyway, these two effects imply the same conclusion: The backreaction of the RSET will doom the warp drive to be semiclassically unstable.

\section{Summary of results}\label{sec:results}

We think that this work is convincingly ruling out the semiclassical stability of superluminal warp drives on the base of the following evidence.

(1) We found that the central region of the warp drive behaves like the asymptotic region of a black hole: In both of these regions the static term $\rho_{st}$ vanishes and the whole energy density is due to the Hawking radiation generated at the black horizon. If one trusts the QI \cite{pfenningford,broeck}, the wall thickness for a warp drive with $v_0\approx c$ would be $\Delta\lesssim 10^2\,L_{\rm P}$, and its surface gravity $\kappa_1 \gtrsim 10^{-2}\,t_{\rm P}^{-1}$, where {  $t_{\rm P}\approx 10^{-43}$~s} is the Planck time. Hence, the Hawking temperature of this radiation would be unacceptably large: $T_{\rm H} \sim \kappa_1\gtrsim 10^{-2}\,T_{\rm P}$.

(2) The formation of a white horizon produces a transient radiation which accumulates on the white horizon itself. This causes the energy density $\rho$, as seen by a free-falling observer, to grow unboundedly with time on this horizon. The semiclassical backreaction of the RSET will make the superluminal warp drive to become rapidly unstable, in a time scale of the order of $1/\kappa_2$ (i.e.~of the inverse of the surface gravity of the white horizon). In fact, in order to get even a time scale $\tau\sim1 $ s for the growing rate of the RSET, one would need a wall as large as $3\times10^8$ m. Thus, most probably, one would be able to maintain a superluminal speed for just a very short interval of time.

(3) The formation of a Cauchy horizon gives rise to an instability, similar to inner horizon instability in black holes, due to the blue-shift of Hawking radiation produced by the black horizon.

\section{Understanding the nature of the postulate}\label{sec:postulatenature}

We have just reported another episode in the search for failures of the light postulate. Once more the postulate came out of this trial triumphant. So a strong formulation of it seems somehow encoded in natural laws. Can this have a deeper meaning? Is it just a limitation to our possibility to travel and communicate or is it required by consistency in the spacetime fabric? As a matter of fact, any mechanism for superluminal travel can be easily turned into a time machine and hence lead to the typical causality paradoxes associated with these mind-boggling solutions of GR. For instance, in~\cite{everett} it was shown that two warp-drive bubbles traveling in opposite directions can be used to generate closed timelike curves (see also~\cite{everett-roman,visser-book} for causality problems with the existence of two Krasnikov tubes and a two-wormhole system, respectively). 

The mainstream opinion in this respect is that generically the physics associated to GR plus QFT (the same theoretical framework we used in our investigation) is always able to avoid the formation of time machines. This is the so called  Hawking's chronology protection conjecture~\cite{hawking-chronology}. Unfortunately, this conjecture is not yet proved given that: (1) we are not yet able at the moment to perform a self-consistent calculation taking into account the RSET back-reaction on a given spacetime, (2) the Kay-Radzikowski-Wald theorem \cite{krw} implies the breakdown of the renormalizability procedure of the SET on chronological horizons (which are just a special sort of Cauchy horizons). See also~\cite{visser-chronology} for an extensive review on the present status of the chronology conjecture. 

The results presented in this essay suggest an interesting twist about the way this conjecture could be enforced in nature. Indeed, it might be that chronology protection is just a side consequence of a strong form of the speed of light postulate. That is, ``spacetime shortcuts'' like warp drives, wormholes and Krasnikov tubes might turn out  to be semi-classically unstable (albeit via different mechanisms) whenever one tries to generate them from approximately flat spacetime. This probably deserves further investigation.

While the previous discussion refers to the standard framework (GR plus QFT), different outcomes for the speed of light postulate can be envisaged when departures from GR are taken into account. The search for such departures has been boosted in recent years by rising of the emergent gravity paradigm. Within this framework it is in fact very natural to see also Lorentz invariance as an emergent spacetime symmetry broken at high energy. Indeed, we have nowadays several toy models where a finite speed of propagation can emerge in systems having no fundamental speed limit~\cite{Barcelo:2005fc}. For example, this is the case with the speed of sound in Newtonian (non-relativistic) condensed matter systems. Individual particles of the system can move at arbitrarily large speeds; however, collective density disturbances of wavelengths larger than the inter-particle distance, all propagate at the same finite speed, the speed of sound. 

If electromagnetic fields were emergent collective excitations of an underlying system with the speed of light playing the role of the speed of sound, then, any particle or excitation moving at speeds large than $c$ would slow down by emitting electromagnetic radiation, much as in the \v{C}erenkov effect. The speed of light will appear as insurmountable in practice. This perspective offers an answer to the question with which we started this essay: {\em Because all the physics that we know of, even that in accelerators, is low-energy physics and all the known fields collective variables of a yet unknown underlying system. Maybe it is allowed to travel faster that light, but only for high-energy beings.}

The breakdown of Lorentz invariance generally manifest itself via dispersion relations for matter modified at energies close to the Planck scale, {about $10^{19}$~GeV}.  In this case one generically expects dramatic modifications 
of the behavior of light rays close to the horizons. This in turn could lead to a taming of the exponential growth of the RSET and a late time stabilization of the warp drive. In this sense one could see the results regarding the stability of white hole horizons in QFT with UV Lorentz violations reported in \cite{Macher:2009tw} as an interesting hint {for} further investigation.

To end up this essay let us comment that the light postulate itself is not such a strong limitation for the exploration of the universe as it might seem. Imagine that at the same Earth's time all its inhabitants were separated into several expeditions prepared to visit different star systems at similar distances from the Earth. All the groups {would} have starships able to reach speeds closer to the speed of light. Then, all people could travel to their chosen star system, explore it during some fixed period and return back to the Earth having spent all of them approximately the same amount of proper time, which could be reasonably short if the attained speeds during the expedition were very close to the speed of light. Therefore, for a nomadic society composed by travelers, the exploration limits {would} not come from the light postulate, but from the maximum attainable accelerations of the starships compatible with our structural resistance. 
But this is another story. Let us just say that, as far as we know, traveling at just 99\% of the speed of light would be not that bad, after all.

\begin{acknowledgments}
\emph{Acknowledgments}.--- The authors wish to thank S.~Sonego and M.~Visser for illuminating discussions. S.F. acknowledge the support provided by a INFN/MICINN collaboration. C.B. has been supported by the Spanish MICINN under project FIS2008-06078-C03-01/FIS and Junta de Andalucia under project FQM2288.
\end{acknowledgments}

\bibliography{warpdrive}

\end{document}